\documentstyle[preprint,eqsecnum,aps]{revtex}

\begin{document}
\draft
\preprint{}
\title{Magnetic Dynamics in Underdoped YBa$_2$Cu$_3$O$_{7-x}$: 
Direct Observation of a Superconducting Gap\\}
\author{P. Dai,$^1$ M. Yethiraj,$^1$ H. A. Mook,$^1$ 
T. B. Lindemer,$^1$ and F. Do$\rm\breve{g}$an,$^2$}
\address{
$^1$Oak Ridge National Laboratory, Oak Ridge, 
Tennessee 37831-6393\\
$^2$Department of Materials Science and Engineering\\ 
University of Washington, Seattle, Washington 98195\\}
\date{\today}
\maketitle
\begin{abstract}
Polarized and unpolarized triple-axis neutron measurements were performed on
an underdoped crystal
of YBa$_2$Cu$_3$O$_{7-x}$ 
($x=0.4\pm0.02$, $T_c=62.7$ K). Our results indicate, contrary to earlier evidence, that the  
spin excitations in the superconducting state are essentially the 
same as those in the fully doped material except that
the  unusual  41 meV resonance is observed at 34.8 meV. 
The normal state spin excitations are characterized by a weakly energy-dependent
continuum whose
temperature dependence shows the 
clear signature of a superconducting gap at $T_c$. 
The enhancement at the resonance is accompanied by a suppression of magnetic 
fluctuations 
at both higher and lower energies.
\end{abstract}
\pacs{PACS numbers: 74.72.Bk, 61.12.Ex}

\narrowtext
Recently, there has been a great deal of interest in
 underdoped superconducting copper oxides
 because these materials display unusual normal state properties \cite{anderson}
 that suggest 
the opening of a pseudogap (a suppression of spectral weight) 
 in the spin or charge excitation spectrum well above $T_c$ \cite{levi}.
The central issue is whether the physics of underdoped cuprates is 
fundamentally different from the description of the  
Bardeen-Cooper-Schrieffer (BCS) theory of superconductivity.
Although there is ample evidence \cite{walstedt,marshall,homes}
for the existence of a pseudogap in underdoped cuprates, no one has
established a direct correlation between the pseudogap and the 
superconductivity.
Furthermore, it is not known whether a superconducting gap opens at 
$T_c$.
  
Magnetic inelastic
neutron scattering  
is a powerful technique for measuring 
the wavevector dependence of the superconducting 
gap function \cite{joynt} because the 
imaginary part of the dynamic susceptibility, 
$\chi^{\prime\prime}({\bf q},\omega)$,
probed by neutrons is directly associated with electron-hole pairs formed by 
transitions across the Fermi surface. 
In the BCS theory of 
superconductivity, such transitions become disallowed for all
wavevectors with energy transfer ($\hbar\omega$) below 2$\Delta$ 
($3.52k_BT_c$ for the weak-coupling limit) at $T=0$ K, where $\Delta$ is the 
superconducting gap.
Since the availability of 
large single crystals of YBa$_2$Cu$_3$O$_{7-x}$ [denoted as (123)O$_{7-x}$], 
there have been a 
large number
of neutron scattering experiments on this system across the whole doping
range \cite{mignod,tranquada,mook,bourges,fong}. 
The nature of the scattering at low temperatures in the 
superconducting state of  highly doped (123)O$_{7-x}$ 
appears established. 
Early measurements by Rossat-Mignod and co-workers \cite{mignod} indicated an 
enhancement of scattering in the superconducting state near 41 meV. 
Polarized neutron experiments by Mook {\it et al.} \cite{mook} have 
shown that below
about 50 meV the magnetic scattering is dominated by 
 a peak that is sharp in energy, located at 41 meV,
and centered at $(\pi,\pi)$ in the reciprocal lattice. 
The peak is generally referred to as a resonance peak and
a number of theoretical explanations have been proposed \cite{theory}. 
Since the resonance disappears in the 
normal state \cite{fong},
 it may be a signature of the superconducting state.

Although there is general agreement about the low temperature 
spin excitations for highly doped (123)O$_{7-x}$, 
$\chi^{\prime\prime}({\bf q},\omega)$ for underdoped compounds 
($0.6>x\geq0.35$) has been far from
clear. Using unpolarized neutrons,
  Rossat-Mignod and co-workers 
\cite{mignod,bourges} as well as Tranquada {\it et al.} \cite{tranquada}
have reported two distinctive features in $\chi^{\prime\prime}({\bf q},\omega)$
that are not present for highly doped (123)O$_{7-x}$. 
The first of these is an energy gap in the 
spin excitation spectrum (below which no magnetic 
intensity can be found) which 
is much less than 2$\Delta$. 
They also reported a complicated temperature dependence
of $\chi^{\prime\prime}({\bf q},\omega)$, but found no
resonance coupled directly to $T_c$.

We report here the first polarized neutron scattering 
measurements on an underdoped single crystal of 
(123)O$_{7-x}$ ($x=0.4\pm0.02$, $T_c= 62.7$ K). These experiments are 
able to isolate the magnetic scattering from nonmagnetic sources such as phonons and   
demonstrate that a clear superconducting gap opens 
in the spin excitation spectra at $T_c$. This gap, observed 
at wavevector $(\pi,\pi)$, is 28 meV (5.2$k_BT_c$) at the lowest 
temperatures which is large compared to that expected in the weak-coupling BCS limit. 
We also show, contrary to earlier evidence, that the low 
temperature spin excitations in the superconducting state are very much 
like those found for the fully doped material except for a change in 
the energy scale. The unusual resonance, found at 41 meV in the 
fully doped materials, is thus not unique to 
materials with high doping. In fact, the resonance energy falls with 
$T_c$ and is at 34.8 meV for (123)O$_{6.6}$.

The polarized and unpolarized beam experiments have been performed 
on a twinned disk shaped crystal (123)O$_{6.6}$ (22.9 mm in diameter and 
10.8 mm in thickness)
 that weighs 25.59-g and has a mosaic of 0.9$^\circ$ as measured from 
the (006) reflection. 
Similar to other large single crystals of (123)O$_{7-x}$ used in 
previous neutron scattering experiments \cite{mook,fong}, our 
sample contains approximately 14 mol\%  
Y$_2$BaCuO$_5$ as an impurity \cite{raveau} determined by
neutron
 powder diffraction and Rietveld analysis on identically 
prepared samples. Oxygenation of large, dense (123)O$_{7-x}$ samples
suitable for neutron scattering experiments is not a trivial 
matter. For example, although the $T_c$ of (123)O$_{7-x}$ exhibits the
well known ``60 K plateau'' for $x\approx 0.4$ 
\cite{munro}, previous neutron scattering samples with the nominal
oxygen content of (123)O$_{6.6}$ \cite{tranquada} have had a 
$T_c$ variation of as much as 7 K. To overcome this 
difficulty, an improved technique for oxygenating dense (123)O$_{7-x}$ was 
used. Briefly, the sample was oxygenated in a thermogravimetric apparatus (TGA)  
by annealing to the equilibrium  weight ($\pm0.1$-mg) at $T$-$p$[O$_2$] conditions 
for (123)O$_{6.6}$ as calculated from the thermodynamic
models that were fitted to experimental data \cite{lin1}.
Instead of quenching the sample after annealing, 
the partial pressure of  oxygen (99.999\% purity) in the TGA 
 was adjusted continuously throughout the
cooling process to maintain a 
constant sample mass, thus ensuring 
uniformity in oxygen content and an equilibrium crystal structure \cite{lin2}.  
The advantage of such technique is clearly evident in the AC susceptibility 
measurement shown in Fig. 1 which
gives a superconducting onset temperature of 62.7 K, the highest 
for any (123)O$_{6.6}$ samples studied by neutron scattering.
The room temperature lattice parameters 
 [$a=3.8316(5)$ \AA, 
$b=3.8802(6)$ \AA, and $c=11.743(2)$ \AA], 
the orthorhombicity [200(b-a)/(b+a)=1.2604], and the unit cell volume 
($V=174.59$ \AA$^{3}$) are consistent with the (123)O$_{6.6}$ 
powder sample oxygenated using the
identical methods \cite{chakoumakos}.

Inelastic measurements were made at the High-Flux Isotope Reactor at
Oak Ridge National Laboratory  
using the HB-1 and HB-3 triple-axis 
spectrometers with a fixed final neutron energy of 30.5 meV
and a pyrolytic graphite (PG) filter before the
analyzer \cite{mook}.
To separate magnetic from nuclear scattering 
in polarized beam experiments, 
it is often sufficient to measure the spin-flip scattering
(SF) and the non-spin-flip (NSF) scattering. In cases where
the magnetic signal is on top of a large background, the SF
scattering can be measured with neutron polarization first 
parallel to the momentum transfer (horizontal guide field or HF) and
then perpendicular (VF). The difference (HF-VF) gives one-half
of the magnetic intensity completely free of background effects \cite{moon}.
 
For the experiment we index reciprocal ($Q$) space by using the orthorhombic unit cell so that 
momentum transfers ($Q_x,Q_y,Q_z$) in units of 
\AA$^{-1}$ are at positions 
(H,K,L)=$(Q_xa/2\pi,Q_yb/2\pi,Q_zc/2\pi)$ reciprocal lattice units (r.l.u.).
Most of our measurements were made in the 
(H,3H,L) Brillouin zone.
For highly doped (123)O$_{7-x}$, 
this scattering 
geometry was first used by Fong {\it et al.} \cite{fong}  
and 
has the advantage of having 
less phonon and accidental Bragg scattering.
	
Since previous investigations 
on highly doped (123)O$_{7-x}$ showed
magnetic scattering that peaked at 
 $(\pi,\pi)$ with a    
sinusoidal modulation along the (0,0,L) direction, 
we looked for magnetic
signal in the normal and superconducting 
states at $(-0.5,-1.5,1.7)$ which corresponds to the 
maximum intensity of the (0,0,L) modulation.							
The results of the measurement are shown in Fig 1. 
The dynamic susceptibility is featureless at 75K, but at 11K is dominated by
a peak at $\sim$34 meV. 
In addition, the low 
energy excitations ($\hbar\omega\leq28$ meV) vanish 
 to within the error 
of the measurements, consistent with the opening of 
an energy gap. The NSF data indicates that
the phonon scattering is 
essentially unchanged in this zone between 11 K and 75 K.
Consequently, our conclusions about the magnetic excitations 
in underdoped (123)O$_{6.6}$ are different from those of 
previous unpolarized experiments \cite{mignod,tranquada,bourges} 
which reported a broad peak
around 22 meV in the normal state and observed no 34 meV resonance or 
28 meV gap in the superconducting state. 

Figure 2 summarizes the wavevector dependence of the magnetic excitations. 
The 24 meV constant-energy scan at 75K peaks at $(\pi,\pi)$ with
a full width half maximum of 0.32 \AA$^{-1}$ which
corresponds to an antiferromagnetic correlation length of 
$\sim$20 \AA .  
The same scan was taken at 11 K with both HF and VF.
To  within the error of the 
measurement, no magnetic intensity is present. 
Similar scans were made at $\hbar\omega=34$ meV. 
The intensity of the resonance peak clearly increases
below $T_c$. 
The NSF phonon scattering (not shown) for these scans are 
featureless and changes negligibly 
between these temperatures.

Since the phonon scattering does not change between
11 K and 75 K in the (H,3H,L) zone, unpolarized neutrons are
used to determine the temperature dependence of the resonance.  
Comparison of the difference spectra in Fig. 3  
 reveals that the resonance softens and broadens 
slightly at high temperatures. 
The energy threshold below which the scattering is suppressed, indicated 
by arrows in the figure, moves 
 from 25 meV at 50 K to 28 meV at 11 K, indicating 
that the gap increases in energy with cooling.
  In addition, the energy width of the 
resonance narrows from 6.6 meV at 50 K to 5.5 meV at 11 K;
and is resolution limited at 11 K. 
The negative values in the 
subtractions of Fig. 3 show the opening 
of an energy gap in $\chi^{\prime\prime}({\bf q},\omega)$, consistent
with polarized results of Fig.1. Furthermore, our data [Fig. 3(c)]
suggest that the enhancement at the resonance is accompanied by
a suppression of the fluctuations both below and above it. This point 
is clear from negative values on both side of the resonance. A detailed
comparison between (123)O$_{6.93}$ and (123)O$_{6.6}$
will be reported in a forthcoming paper \cite{dai}.

One signature of the magnetic fluctuations in (123)O$_{7-x}$ is its
sinusoidal intensity modulation along the (0,0,L) direction. 
These modulations, seen in all the (123)O$_{7-x}$ 
neutron scattering experiments  
\cite{mignod,tranquada,mook,bourges,fong}, are the result of 
the antiferromagnetic   
CuO$_2$ bilayer coupling and may have important theoretical 
consequences \cite{theory,sudip}. Since there are no magnetic fluctuations at 
low temperatures below 28 meV, the difference 
spectra between 75 and 11 K should give an accurate account of the normal
state magnetic scattering. Figure 3(d) shows such a scan along the L 
direction at $\hbar\omega=24$ meV 
together with a calculated profile for the acoustic mode of
the antiferromagnetic bilayer coupling \cite{mignod,tranquada}. Clearly, 
the bilayer modulation is present in the normal state and
the acoustic mode adequately describes the data.

In Figure 4 we plot the temperature dependence of 
scattered intensity at 24 and 
35 meV which corresponds to frequencies below and above 
the gap energy, respectively.  
The intensity for both frequencies changes
drastically at $T_c$, but in opposite directions. 
Below $T_c$, the intensity at $\hbar\omega=24$ meV 
decreases precipitously
showing the opening of the superconducting gap 
while the temperature
dependence of the resonance at 35 meV parallels that of the 41 meV resonance 
for highly doped (123)O$_{7-x}$ \cite{mook,bourges}. 
Since the polarized HF$-$VF data at 24 meV 
shows no magnetic signal at 11 K, 
we are able to determine the low temperature background (see Fig 4).
The corresponding $\chi^{\prime\prime}({\bf q},\omega)$ can then be calculated
via the fluctuation-dissipation theorem and shows a clear suppression
at $T_c$. An unusual feature of the temperature dependence of the 
resonance is that although a large increase in intensity takes place below
$T_c$,  a smaller increase is observed that extends to temperatures well
above $T_c$.
No such behavior is observed for ideally doped (123)O$_{6.93}$ \cite{mook,bourges}.

In conclusion, we have shown that magnetic excitations for underdoped 
(123)O$_{6.6}$ are essentially the same as those in the fully doped 
material except for a change in the energy scale. The normal state scattering
is characterized by a weakly energy-dependent continuum whose temperature
dependence shows the clear signature of a superconducting gap at $T_c$.
      
We thank G. Aeppli, P. W. Anderson, D. Pines, D. J. Scalapino, D. G. Mandrus, 
G. D. Mahan, B. C. Chakoumakos, R. F. Wood, and J. Zhang for helpful discussions. 
This research was supported by the US DOE under 
Contract No. DE-AC05-96OR22464 with Lockheed Martin 
Energy Research, Inc.

\begin{figure}
\caption{(a) 
Polarized beam measurements of 
$\chi^{\prime\prime}({\bf q},\omega)$ for the
$(-0.5,-1.5,1.7)$ r.l.u. position at 75 K and (b) 11 K, 
obtained by subtracting the
background [19], correcting 
for incomplete polarization, and dividing by the 
Bose population factor $n(\omega)$+1. 
The filled circles represent twice 
the SF, HF-VF, scattering. The flipping ratio for HF and VF are 
$15.8\pm0.25$ and $16.5\pm0.26$, respectively. The solid lines are
guides to the eye. 
(c) The phonon scattering normalized to 600 
monitor counts.  
(d) AC susceptibility of the sample, showing the
onset $T_c$ of 62.7 K with a transition width that is 3.3 K wide.}
\label{autonum}
\end{figure}

\begin{figure}
\caption{(a) Polarized SF, (H,3H,1.7) scans using HF 
for $\hbar\omega=24$ meV at $T=75$ K, and (b) 11 K. 
The filled circles in (b) are the
SF scattering using VF. HF-VF scattering shows no magnetic 
signal.
(c) SF, HF, (H,3H,1.7) scan for $\hbar\omega=34$ meV 
at $T=75$, and (d) 11 K. 
Solid lines in (a), (c), and 
(d) are Gaussian fits to the data. The solid line in (b) is
a guide to the eye.}
\end{figure}

\begin{figure}
\caption{Unpolarized beam measurements at (0.5,1.5,1.7) r.l.u. in 
which the data at 75 K are subtracted 
from (a) 50 K, (b) 30 K, and (c) 11 K data. For energy transfers 
above 45 meV, the data was taken with neutron final energy of
35 meV. The lines are fits to
Gaussians and linear backgrounds. Arrows in the figure indicate
 points defined arbitrarily at $1\%$ of the resonance
intensity. (d) Unpolarized beam constant-energy 
scan along (0.5,1.5,L) at $\hbar\omega=24$ meV 
 using HB-1. The spectrum is obtained by subtracting the 
signal at 75 K from the $T=11$ K background. The solid line is
the expression $f_{Cu}^2({\bf Q})\sin^2(\pi z L)$ normalized to
the intensity at $(0.5,1.5,-1.7)$, where
$f_{Cu}({\bf Q})$ is the Cu magnetic form factor, and $z\cdot c = 3.342$ is the
distance between adjacent copper-oxide planes.}
\end{figure}

\begin{figure}
\caption{Temperature dependence of the scattering at (0.5,1.5,1.7) for
(a) $\hbar\omega=24$, and (b) 35 meV. (c) $\chi^{\prime\prime}({\bf q},\omega)$ 
at $\hbar\omega=24$ meV obtained by 
subtracting the nonmagnetic scattering and dividing by the Bose
population factor $n(\omega)+1$. (d) Temperature dependence of
the polarized SF scattering at $(-0.5,-1.5,1.7)$ for $\hbar\omega=34$ meV.
The filled circles represent the peak intensity of constant-energy scans and
Gaussians fits to the data (Fig. 2) show a temperature independent
 background of $20\pm1.5$ counts.
Arrows in the figure indicate the onset of $T_c$ and solid lines are
guides to the eye.}
\end{figure}

\end{document}